\input harvmac.tex
\vskip 2in
\Title{\vbox{\baselineskip12pt
\hbox to \hsize{\hfill}
\hbox to \hsize{\hfill HIP-2001-61/TH}}}
{\vbox{\centerline{BRST properties of New Superstring States}
\vskip 0.3in
{\vbox{\centerline{}}}}}
\centerline{Dimitri Polyakov\footnote{$^\dagger$}
{polyakov@mappi.helsinki.fi}}
\medskip
\centerline{\it Dept. of Physical Sciences}
\centerline{\it Helsinki University}
\centerline{\it and Helsinki Institute of Physics}
\centerline{\it PL 64 }
\centerline{\it FIN-00014  Helsinki, Finland}
\vskip .5in
\centerline {\bf Abstract}
Brane-like states are defined  by physical vertex operators in NSR
superstring theory, existing at nonzero pictures only.
These states exist both in open and closed string theories,
in the NS and NS-NS sectors respectively.
In this paper we present a detailed analysis of their BRST properties,
giving a proof that these vertex operators are physical,
i.e. BRST invariant and BRST non-trivial.
\Date{November 2001}
\vfill\eject
\lref\myself{D.Polyakov, hep-th/9812044,Phys.Lett. {bf B470:103-111}}
\lref\myselff{D.Polyakov, hep-th/0005094, Class.Quant.Grav. {bf
18:1979, 2001}}
\lref\fms{D.Friedan, E.Martinec, S.Shenker, Nucl.Phys.{\bf B271:93,1986}}
\lref\azc{
J.A. de Azcarraga, J.P. Gauntlett, J.M. Izquierdo, P.K. Townsend,
Phys.Rev.Lett.62:2579-2582,1989}
\lref\kts{R.Kallosh, A.Tseytlin,{\bf J.High Energy Phys.9810:016,1998}}
\lref\ampf{S.Gubser,I.Klebanov, A.M.Polyakov,
Phys.Lett. {\bf B428:105-114}}
\lref\pol{ J. Polchinski, Phys.Rev.Lett,
 hep-th/9901076 }
\lref\malda{J.Maldacena, Adv.Theor.Math.Phys.2 (1998)
231-252, hep-th/9711200}
\lref\wit{E.Witten{\bf Adv.Theor.Math.Phys.2:253-291,1998}}
\lref\kleb{I.R. Klebanov, A.M. Polyakov, Mod.Phys.Lett.A6:3273-3281,1991}
\centerline{\bf Introduction}
Critical superstring theory in the NSR formalism contains physical
vertex operators
~refs{\myselff, \myself}
with ghost-matter mixing, i.e. the operators that exist at non-zero
pictures only.
These vertex operators do not correspond to any point-like particle in
the perturbative superstring spectrum but instead are closely related to
the dynamics of
extended objects, such as D-branes. These operators are therefore expected
 to play a crucial role
in non-perturbative physics of strings and somehow it appears that it is
 an essential mixing of
matter with superconformal ghosts in these vertex operators that  is
relevant to their non-perturbative nature.
In the open string case, these massless vertex operators, one two-form
(at picture -2) and two five-forms
(at pictures $-3$ and $+1$ ), are given by:
\eqn\grav{\eqalign{V_5^{(-3)}=\oint{{dz}\over{2i\pi}}{e^{-3\phi}}
\psi_{m_1}...\psi_{m_5}e^{ik^{\perp}X}(z)\cr
V_5^{(+1)}=\oint{{dz}\over{2i\pi}}
e^{\phi}\psi_{m_1}...\psi_{m_5}e^{ik^{\perp}X}+ghosts\cr
V_2^{(-2)}=\oint{{dz}\over{2i\pi}}
e^{-2\phi}\psi_{m_1}\psi_{m_2}e^{ik^{\perp}X}}}
Bosonization expressions for superconformal ghosts are the
standard ones; see the formula (88) of this paper.
The contour integrals in (1) are taken over the
open string worldsheet boundary; $\psi$'s are NSR worldsheet fermions and
 $\phi$ is bosonized superconformal ghost field; the space-time indices
$m_i$ run from 0 to 9. The picture $+1$ five-form vertex also contains
fermionic ghost part (consisting of two terms with  fermionic
ghost numbers $+1$ and $-1$);
these fermionic ghost terms are necessary to insure the BRST
invariance of this vertex operator;
in this paper we will particularly give a detailed derivation of
this fermionic ghost part.
One important property of the
brane-like states distinguishing them from usual vertex operators
in ten-dimensional string theory, such as a photon or a graviton,
 is their propagation: namely,  as we will show in details in Section 2
of this paper,
the BRST invariance and non-triviality conditions
altogether restrict their propagation to lower dimensional subspaces;
 that is, their
 momenta $k^{\perp}$ must be  taken $transverse$ with respect to their
$m_i$ space-time indices.
Thus $V_5^{(+1)}$ and $V_5^{(-3)}$
 are constrained to propagate in five-dimensional
subspace of $R^{10}$ while
the two-form is allowed to propagate in eight dimensions.
In this respect, the brane-like states (1)
are reminiscent of a multi-dimensional
analogue of the discrete states in two-dimensional gravity ~{\kleb}.
The  closed string brane-like states may be
constructed from the open string ones
by multiplying them by a suitable anti-holomorphic
 part; there are in fact various ways of doing
that corresponding to  various brane configurations.
As usual, the vertex operators for the open-string brane-like states
may also be written in an
alternative, unintegrated form:

\eqn\grav{\eqalign{V_5^{(-3)}=c{e^{-3\phi}}\psi_{m_1}...\psi_{m_5}
e^{ik^{\perp}X}(z)\cr
V_5^{(+1)}=
ce^{\phi}\psi_{m_1}...\psi_{m_5}e^{ik^{\perp}X}+ghosts\cr
V_2^{(-2)}=
ce^{-2\phi}\psi_{m_1}\psi_{m_2}e^{ik^{\perp}X}}}
where $c$ is  fermionic  ghost field.

In this paper we perform the detailed analysis of the BRST properties of
the brane-like vertex operators
proving that they correspond to well-defined physical states in  the NSR
This paper is organized as follows.
In the first section we review arguments relating
the brane-like states to
D-brane dynamics. We show that zero momentum parts of these operators
appear as central terms in picture-changed superalgebras,
thus corresponding to brane topological charges.
We also demonstrate that these
vertex may be considered as a special
limit  of the RR states in non-canonical pictures.
In the sections 2 and 3 we  analyze BRST properties of the brane-like
states. In the section 2 we prove the BRST invariance
of the brane-like states (1),(2). While for the two-form
and the 5-form at the picture $-3$ the proof is elemenrtary
and straightforward, to insure the BRST invariance of the
picture $+1$ five-form one has to introduce two fermionic ghost
counterterms, carrying fermionic ghost numbers $+1$ and $-1$.
 We give the detailed derivation of these counterterms
and show that their ghost structure is similar to the
one present in the BRST invariant expression for picture-changing
operator ~{\fms}
 Next, in the section 3 we give the proof that the states (1), (2),
 are BRST-nontrivial, i.e. cannot be represented as a
BRST commutator with any operator in a small Hilbert space.
At the same time, the BRST non-triviality condition
will be shown to impose significant constraints
on propagation of the 5-form state. Unlike
the case of conventional perturbative vertex operators in
critical superstring
theory propagating in 10 dimensions, such as a graviton or a photon,
the five-form state will be shown to be constrained to propagate in
the five-dimensional subspace, transverse to the one spanned by the
$m_1$,...$m_5$ indices.
Thus in this paper we give complete and detailed
proof that the new superstring states (1)  are
physical and therefore must play an important role in
non-perturbative superstring
dynamics.
 In the concluding section we construct closed-string version
of the brane-like state by a straightforward generalization
of an open string case, and also
present some new results on the contribution of the
brane-like states (1) to the low-energy effective action
of superstring theory to demonstrate that it reproduces
the DBI expression for D-brane actions.

\centerline{\bf Brane-like States and Central Terms in
space-time SUSY algebras}

In this section we will show that
 the vertices (1) at zero momentum
appear as central terms in picture-changed superalgebras
in the presence of D-branes, which points
at their relation to brane dynamics.
Consider the space-time supercharge at the  canonical
$-1/2$-picture:
\eqn\grav{\eqalign{Q^{-1/2}_\alpha=\oint{{dz}\over{2i\pi}}
e^{-{\phi\over2}}\Sigma_\alpha(z)}}
where $\alpha,\beta,...$ denote 10d spinor indices.
It is easy to check that this supercharge satisfies the standard SUSY
algebra:
\eqn\grav{\eqalign{
\lbrace{Q^{-1/2}_\alpha;Q^{-1/2}_\beta}\rbrace=
\gamma^m_{\alpha\beta}P_m^{(-1)}}}
where
\eqn\lowen{P_m^{(-1)}=\oint{{dz}\over{2i\pi}}e^{-{\phi}}\psi_m}
is a translation operator at the $-1$-picture.
Consider, however, a deformation of this supercharge by
a BRST exact space-time spinor,
$T_\alpha={e^{-{{3\phi}\over2}}\Sigma_\alpha(z)}$
Note that even though $T_\alpha$ is BRST exact, its
propagation is described by a
physical state $T_\alpha{e^{ikX}}$ which is BRST non-trivial
at nonzero momentum.
In particular, it is well-known that the
Ramond-Ramond physical state, $T{\bar{S}}e^{ikX}$ is
the source of the RR gauge potential $A_{RR}(k)$,
rather than the RR field strength (unlike the canonical picture case).
Denoting the deformed space-time SUSY
charge as $S_\alpha=Q_\alpha+T_\alpha$,
one can easily show that it satisfies the superalgebra:
\eqn\grav{\eqalign{\lbrace{S_\alpha;S_\beta}\rbrace=
\gamma^m_{\alpha\beta}P_m^{(-1)}+
\gamma^{m_1m_2}_{\alpha\beta}Z_{m_1m_2}
+\gamma^{m_1...m_5}_{\alpha\beta}Z_{m_1...m_5}}}
where
the 2-form and the 5-form central terms are given by
\eqn\grav{\eqalign{Z_{m_1m_2}=\oint{{dz}\over{2i\pi}}
e^{-2\phi}\psi_{m_1}\psi_{m_2}\cr
Z_{m_1...m_5}=\oint{{dz}\over{2i\pi}}
e^{-3\phi}\psi_{m_1}...\psi_{m_5}}}
i.e. they are given by the zero momentum parts
of the vertices (1).
It is well-known that p-form central terms in SUSY algebra
always correspond to topological charges of
p-branes, i.e. non-perturbative extended objects.
Therefore we expect that the vertex operators
(1), if they are physical, must describe the non-perturbative
brane dynamics.
The non-perturbative nature of these operators should
be related to the fact that they don't exist at zero picture,
i.e. to the ghost-matter mixing.

Because of the topological nature of the central charges (7),
and because of the BRST triviality of $T_\alpha$ one should
expect that the operators (1)  are BRST exact at zero momentum,
however they should be physical at nonzero $k$,
just like in the case of picture $-3/2$ space-time spinor
$T_\alpha{e^{ikX}}$. In the sections 2 and 3 of the paper
we will prove that this  is precisely the case,
i.e. the operators (1) at $k\neq{0}$ indeed define
physical superstring excitations.
Finally, the vertices (1) may be understood as a special
singular limit of Ramond-Ramond states at
non-canonical pictures.
Indeed, consider
the RR vertices:
\eqn\grav{\eqalign{V_{RR}^{-3/2,-3/2}=
T_\alpha{\bar{T}}_\beta{e^{ikX}}(z,{\bar{z}})\cr
V_{RR}^{-3/2,-1/2}=T_\alpha{\bar{S}}_\beta{e^{ikX}}(z,{\bar{z}})
}}
(of course they should be multiplied by appropriate combinations
of 10 d gamma-matrices and Ramond-Ramond tensor fields).
Consider these operators placed on the disc,
i.e. in some D-brane background.
Because of the disc boundaries, the holomorphic and antiholomorphic
parts of these operators are no longer independent but
have a non-trivial O.P.E. product.
Whenever any of these operators approaches the edge of the D-brane,
or the disc boundary, this O.P.E. becomes singular, proportional to
${1\over{z-{\bar{z}}}}$ multiplied by some boundary operator.
It is easy to check (as the calculation is totally similar to
computing the anticommutators in the SUSY algebra (6))
that this boundary operator is given exactly by
the two-form $Z_{m_1n_2}e^{ikX}$ in case of
the $V_{RR}^{-3/2,-1/2}$ RR operator, and by the five-form
 $Z_{m_1...m_5}e^{ikX}$ in case of $V_{RR}^{-3/2,-3/2}$,
multiplied by suitable products of 10d gamma-matrices.
Thus the appearance of the operators (1) can be understood as a result of
a normal
reordering inside the Ramond-Ramond vertices  taking place
when the RR charge sources are located near the
D-brane edge. In particular, this is an example of an
 interesting connection between D-brane
Ramond-Ramond and topological charges.

\centerline{\bf 2. BRST-invariance of the Brane-like States}

In this section, we demonstrate the BRST invariance of
the brane-like vertex operators as a first necessary step to show that
they define physical string-theoretic states.
The BRST operator is given by:
\eqn\grav{\eqalign{Q_{BRST}=\oint{{dz}\over{2i\pi}}(c(T_{matter}+T_{ghost})
+{1\over2}e^{\phi-\chi}\psi_m\partial{X^m}+
{1\over4}e^{2\phi-2\chi}b
-b:c\partial{c}:)}}
Let us first of all analyze the question of the BRST invariance
of the vertices (1),(2).
Let us consider the integrated vertices (1).
It is easy to see that the
$V_5^{(-3)}$ and $V_2^{(-2)}$ operators, i.e.
 the picture -3 five-form and the two-form are BRST-invariant.
indeed, since the integrands of $V_5^{(-3)}$ and $V_2^{(-2)}$ are  primary
 fields of
conformal dimension 1, we have
\eqn\grav{\eqalign{{\lbrace}\oint{{dz}\over{2i\pi}}
c(T_{matter}+T_{ghost}(z),V_5^{(-3)}\rbrace
\cr=\oint{{dw}\over{2i\pi}}\lbrack
\oint{{dz}\over{2i\pi}}{1\over{(z-w)^2}}:
c{e^{-3\phi}}\psi_{m_1}...\psi_{m_5}e^{ik^{\perp}X}:(w)\cr+
{1\over{(z-w)}}(:{\partial}c{e^{-3\phi}}
\psi_{m_1}...\psi_{m_5}e^{ik^{\perp}X}:(w)
+:c{\partial}({e^{-3\phi}}\psi_{m_1}...\psi_{m_5}e^{ik^{\perp}X}:(w))\rbrack
\cr=\oint{{dw}\over{2i\pi}}
({\partial}c){e^{-3\phi}}\psi_{m_1}...\psi_{m_5}e^{ik^{\perp}X}
+c{\partial}({e^{-3\phi}}\psi_{m_1}...\psi_{m_5}e^{ik^{\perp}X})\cr=
\oint{{dw}\over{2i\pi}}{\partial}(c{e^{-3\phi}}
\psi_{m_1}...\psi_{m_5}e^{ik^{\perp}X})=0}}
Next, using the O.P.E's:
\eqn\grav{\eqalign{:e^{\alpha\phi}:(z):e^{\beta\phi}:(w)\sim
(z-w)^{\alpha\beta}{:e^{(\alpha+\beta)\phi}:}(w)\cr
:e^{\alpha\chi}:(z):e^{\beta\chi}:(w)\sim
(z-w)^{-\alpha\beta}{:e^{(\alpha+\beta)\chi}:}(w)\
b(z)c(w)\sim{1\over{z-w}}+...\cr
c(z)c(w))\sim(z-w):c\partial{c}:(w)+...\cr
b(z)b(w))\sim(z-w):b\partial{b}:(w)+...
}}
it is easy to see that other terms in the BRST current have no poles in the
 O.P.E.
with ${e^{-3\phi}}\psi_{m_1}...\psi_{m_5}e^{ik^{\perp}X}(w)$ and therefore
\eqn\lowen{\lbrace
\oint{{dz}\over{2i\pi}}({1\over2}e^{\phi-\chi}\psi_m\partial{X^m}+
{1\over4}e^{2\phi-2\chi}b
-b:c\partial{c}:),V_5^{(-3)}\rbrace=0}
and hence
\eqn\lowen{\lbrace{Q_{BRST}},V_5^{(-3)}\rbrace=0}
The BRST invariance of the two-form $V_5^{(-3)}$ is proven totally analogously.
However,  for the picture $+1$ five-form things are more subtle.
The straightforward generalization, simply by replacing the $e^{-3\phi}$ of
$V_5^{(-3)}$
with the $e^{\phi}$ field having the superconformal ghost number $+1$ and
 of the same conformal
dimension $-{3\over2}$, is not BRST invariant, since
the ${e^{-3\phi}}\psi_{m_1}...\psi_{m_5}e^{ik^{\perp}X}:(w)$ operator does
 not commute with
the supercurrent part of the BRST charge given by the terms
 ${1\over2}e^{\phi-\chi}\psi_m\partial{X^m}$ and ${1\over4}e^{2\phi-2\chi}b$
of $Q_{BRST}$.
Namely, calculating the terms giving the simple poles in the relevant
O.P.E's  and evaluating the residue,
one easily gets
\eqn\grav{\eqalign{\lbrace\oint{{dz}\over{2i\pi}}e^{\phi-\chi}\psi_m\partial
{X^m}(z),
{e^{\phi}}\psi_{m_1}...\psi_{m_5}e^{ik^{\perp}X}:(w)\rbrace\cr=
{e^{2\phi-\chi}}\lbrace\psi_{m_1}...\psi_{m_5}(\psi\partial{X})^{\perp}+
\psi_{{\lbrack}m_1}...\psi_{m_4}\partial{X_{m_5\rbrack}}(\partial\phi-
\partial\chi)
\cr+\psi_{{\lbrack}m_1}...\psi_{m_4}\partial^2{X_{m_5\rbrack}}+
i\psi_{m_1}...\psi_{m_5}((k^{\perp}\psi)(\partial\phi-\partial\chi)+
(k^{\perp}\partial\psi))\rbrace{e^{ik^{\perp}X}}:(w)}}
and
\eqn\grav{\eqalign{\oint{{dz}\over{2i\pi}}{\lbrace}
e^{2\phi-2\chi}b(z),{e^{\phi}}\psi_{m_1}...\psi_{m_5}e^{ik^{\perp}X}(w)
\rbrace\cr=
b{e^{3\phi-2\chi}}\psi_{m_1}...\psi_{m_5}(2\partial\phi+2\partial\chi
-\partial\sigma)
{e^{ik^{\perp}X}}(w)}}
therefore
\eqn\grav{\eqalign{{\lbrace}Q_{BRST},{e^{\phi}}\psi_{m_1}...\psi_{m_5}
e^{ik^{\perp}X}:(w)\rbrace\cr=
{1\over2}{e^{2\phi-\chi}}\lbrace\psi_{m_1}...\psi_{m_5}
(\psi\partial{X})^{\perp}+
\psi_{{\lbrack}m_1}...\psi_{m_4}\partial{X_{m_5\rbrack}}(\partial\phi
-\partial\chi)
\cr+\psi_{{\lbrack}m_1}...\psi_{m_4}\partial^2{X_{m_5\rbrack}}+
i\psi_{m_1}...\psi_{m_5}((k^{\perp}\psi)(\partial\phi-\partial\chi)+
(k^{\perp}\partial\psi))\rbrace{e^{ik^{\perp}X}}:(w)\cr
+{1\over4}b{e^{3\phi-2\chi}}\psi_{m_1}...\psi_{m_5}(2\partial\phi
+2\partial\chi-\partial\sigma)
{e^{ik^{\perp}X}}(w)}}
Here and anywhere else $\lbrack{m_1...m_5}\rbrack$
denotes antisymmetrization over the the $m_1,...,m_5$ space-time indices.
Therefore the naive expression for the picture $+1$ five-form, given by
${e^{\phi}}\psi_{m_1}...\psi_{m_5}e^{ik^{\perp}X}(w)$ is not BRST-invariant.
However, below we will show that BRST invariance is restores if one adds
two counterterms to this operators, carrying fermionic ghost numbers
$+1$ and $-1$, so that the full expression for the BRST-invariant
$V_5^{(+1)}$ has the ghost structure similar to the expression
for the picture-changing operator.
Our strategy to find the appropriate counterterms is the following.
Writing schematically
\eqn\grav{\eqalign{\lbrack\oint{{dz}\over{2i\pi}}e^{\phi-\chi}
\psi_m\partial{X^m}(z),
{e^{\phi}}\psi_{m_1}...\psi_{m_5}e^{ik^{\perp}X}:(w)\rbrack=A_1(w)\cr
\lbrack\oint{{dz}\over{2i\pi}}e^{2\phi-2\chi}b(z),{e^{\phi}}\psi_{m_1}...
\psi_{m_5}e^{ik^{\perp}X}(w)\rbrack=A_2(w)}}
we shall aim to find the counterterms $C_1$  and  $C_2$
such that
\eqn\grav{\eqalign{\lbrack\oint{{dz}\over{2i\pi}}e^{\phi-\chi}\psi_m
\partial{X^m}(z),C_1(w)\rbrack
=-{1\over2}A_2(w)\cr
\lbrack\oint{{dz}\over{2i\pi}}e^{2\phi-2\chi}b(z),C_2(w)\rbrack=-2A_1(w)}}
and such that at the same time both $C_1$ and $C_2$ commute with all
other terms
of $Q_{BRST}$ (such as the energy momentum terms and the "opposite" respective
supercurrent term).
Then it is easy to see that the combination
$V_5^{(+1)}={e^{\phi}}\psi_{m_1}...\psi_{m_5}e^{ik^{\perp}X}(w)
+C_1(w)+C_2(w)$ will
 be BRST invariant.
It is also clear that the $C_1$ operator must contain the ghost
 factor of the type  $\sim be^{2\phi-\chi}(w)$ while
the  $C_2$-operator must be proportional to
the factor of the type  $\sim ce^{\chi}(w)$, in other words
the BRST invariant vertex operator
$V_5^{(+1)}$ should have the ghost structure
analogous to that of the picture-changing operator.
Let us start with the derivation of $C_2$.
The ansatz we propose is given by:
\eqn\grav{\eqalign{C_2(w)=-2{\hat{b_3}}c\partial{c}{e^\chi}\partial\chi
(\psi_{m_1}...\psi_{m_5}(\psi\partial{X})^{\perp}+
\psi_{{\lbrack}m_1}...\psi_{m_4}\partial{X_{m_5\rbrack}}
(\partial\phi-\partial\chi)
\cr+\psi_{{\lbrack}m_1}...\psi_{m_4}\partial^2{X_{m_5\rbrack}}+
i\psi_{m_1}...\psi_{m_5}((k^{\perp}\psi)(\partial\phi-\partial\chi)+
(k^{\perp}\partial\psi))){e^{ik^{\perp}X}}(w)}}
and the action of the ${\hat{b}}_n$-operator
on any operator $A$ located at a point $w$
is defined as
\eqn\lowen{{{{\hat{b}}}_n}A(w)=lim_{u\rightarrow{w}}
\lbrack\oint{{dz}\over{2i\pi}}(z-u)^{n+1}b(z),A(w)\rbrack}
If the $A$-operator is located at zero point $w=0$,
then ${{{\hat{b}}}_n}=b_n$
is just the nth oscillation mode of the b ghost field;
while at an arbitrary point $w$ ${{\hat{b}}}_n$ is related to
$b_n$ by appropriate conformal transformation.
Let us now evaluate the commutator of
 $Q_{BRST}$ with $C_2$.
Let us start with the commutator of  $C_2$ with the ghost supercurrent term of
$Q_{BRST}$ given by
${1\over4}\oint{{dz}\over{2i\pi}}e^{2\phi-2\chi}b(z)$.
First of all, since there are no singularities in the O.P.E of b with itself,
it is easy to see that ${\lbrace}\oint{{dz}\over{2i\pi}}
e^{2\phi-2\chi}b(z),{\hat{b}}_3\rbrace=0$ and therefore
\eqn\grav{\eqalign{
{\lbrack}\oint{{dz}\over{2i\pi}}e^{2\phi-2\chi}b(z),C_2(w){\rbrack}\cr=
-2{\hat{b_3}}{\lbrace}\oint{{dz}\over{2i\pi}}e^{2\phi-2\chi}b(z),c\partial{c}
{e^\chi}\partial\chi
(\psi_{m_1}...\psi_{m_5}(\psi\partial{X})^{\perp}+
\psi_{{\lbrack}m_1}...\psi_{m_4}\partial{X_{m_5\rbrack}}(\partial\phi
-\partial\chi)
\cr+\psi_{{\lbrack}m_1}...\psi_{m_4}\partial^2{X_{m_5\rbrack}}+
i\psi_{m_1}...\psi_{m_5}((k^{\perp}\psi)(\partial\phi-\partial\chi)+
(k^{\perp}\partial\psi))){e^{ik^{\perp}X}}(w)}}
Then,
\eqn\grav{\eqalign{L_1\equiv\cr
\equiv{\lbrace}\oint{{dz}\over{2i\pi}}e^{2\phi-2\chi}b(z);
{c}\partial{c}
{e^\chi}\partial\chi
(\psi_{m_1}...\psi_{m_5}(\psi\partial{X})^{\perp}+
\psi_{{\lbrack}m_1}...\psi_{m_4}\partial{X_{m_5\rbrack}}(\partial\phi
-\partial\chi)
\cr+\psi_{{\lbrack}m_1}...\psi_{m_4}\partial^2{X_{m_5\rbrack}}+
i\psi_{m_1}...\psi_{m_5}((k^{\perp}\psi)(\partial\phi-\partial\chi)+
(k^{\perp}\partial\psi))){e^{ik^{\perp}X}}(w)\cr=
(2P^{4}_{2\phi-2\chi-\sigma}c+P^{3}_{2\phi-2\chi-\sigma}(2{\partial}c
+c\partial\chi))
e^{2\phi-\chi}\cr\times
(\psi_{m_1}...\psi_{m_5}(\psi\partial{X})^{\perp}+
\psi_{{\lbrack}m_1}...\psi_{m_4}\partial{X_{m_5\rbrack}}(\partial\phi
-\partial\chi)
\cr+\psi_{{\lbrack}m_1}...\psi_{m_4}\partial^2{X_{m_5\rbrack}}+
i\psi_{m_1}...\psi_{m_5}((k^{\perp}\psi)(\partial\phi-\partial\chi)+
(k^{\perp}\partial\psi))){e^{ik^{\perp}X}}(w)\rbrace}}
where the polynomials
$P^{n}_{\alpha\phi+\beta\chi+\gamma\sigma}$ are defined as:
\eqn\lowen{e^{\alpha\phi+\beta\chi+\gamma\sigma}(z)=
e^{\alpha\phi+\beta\chi+\gamma\sigma}(w)(1+\sum_{n=1}^{\infty}(z-w)^n
P^{n}_{\alpha\phi+\beta\chi+\gamma\sigma}(w))}
for any numbers $\alpha$, $\beta$, $\gamma$
and in particular
\eqn\grav{\eqalign{
P^{4}_{2\phi-2\chi-\sigma}=
{1\over{24}}\lbrace(2\partial\phi-2\partial\chi-\partial\sigma)^4+
(2\partial\phi-2\partial\chi-\partial\sigma)^{\prime\prime\prime}\cr
+4(2\partial\phi-2\partial\chi-\partial\sigma)
(2\partial\phi-2\partial\chi-\partial\sigma)^{\prime\prime}
+3((2\partial\phi-2\partial\chi-\partial\sigma)^{\prime})^2
\cr+6(2\partial\phi-2\partial\chi-\partial\sigma)^{2}
(2\partial\phi-2\partial\chi-\partial\sigma)^{\prime}\rbrace\cr
P^{3}_{2\phi-2\chi-\sigma}={1\over{6}}\lbrace
3(2\partial\phi-2\partial\chi-\partial\sigma)
(2\partial\phi-2\partial\chi-\partial\sigma)^{\prime}\cr
+(2\partial\phi-2\partial\chi-\partial\sigma)^3+
(2\partial\phi-2\partial\chi-\partial\sigma)^{\prime\prime}\rbrace}}
Finally, to compute the commutator of ${1\over4}\oint{{dz}\over{2i\pi}}
e^{2\phi-2\chi}b(z)$
with  $C_2$ we need to act  on $L_1$ with ${\hat{b_3}}$.
Using the O.P.E's:
\eqn\grav{\eqalign{b(z)c(w)\sim{1\over{z-w}}+...\cr
b(z)\partial\sigma(w)\sim{{b(w)}\over{z-w}}+...\cr
b(z)\partial^2\sigma(w)\sim{{b(w)}\over{(z-w)^2}}+...\cr
b(z)(\partial\sigma(w))^2\sim{{b(w)}\over{(z-w)^2}}+...\cr
...\cr
b(z)\partial^4\sigma(w)\sim{{6b(w)}\over{(z-w)^4}}+...\cr
b(z)(\partial\sigma(w))^4\sim{{b(w)}\over{(z-w)^4}}+...}}
 we obtain:
\eqn\grav{\eqalign{{{\hat{b}}_3}L_1(w)={1\over2}
e^{2\phi-\chi}
(\psi_{m_1}...\psi_{m_5}(\psi\partial{X})^{\perp}+
\psi_{{\lbrack}m_1}...\psi_{m_4}\partial{X_{m_5\rbrack}}(\partial\phi
-\partial\chi)
\cr+\psi_{{\lbrack}m_1}...\psi_{m_4}\partial^2{X_{m_5\rbrack}}+
i\psi_{m_1}...\psi_{m_5}((k^{\perp}\psi)(\partial\phi-\partial\chi)+
(k^{\perp}\partial\psi))){e^{ik^{\perp}X}}(w)\cr\equiv{1\over2}
\lbrack\oint{{dz}\over{2i\pi}}e^{\phi-\chi}\psi_m\partial{X^m}(z);
{e^{\phi}}\psi_{m_1}...\psi_{m_5}e^{ik^{\perp}X}:(w)\rbrack}}
Now we will show that
 $C_2$ does commute with all other terms of
$Q_{BRST}$. Indeed, let us consider first the commutator of  $C_2$
with $\oint{{dz}\over{2i\pi}}e^{\phi-\chi}\psi_m\partial{X^m}(z)$.
Again, as before,
since
\eqn\lowen{\lbrace\oint{{dz}\over{2i\pi}}e^{\phi-\chi}\psi_m\partial{X^m}(z),
{{\hat{b}}_3}\rbrace=0}
we write
\eqn\grav{\eqalign{{\lbrack}\oint{{dz}\over{2i\pi}}e^{\phi-\chi}\psi_m
\partial{X^m}(z),C_2(w){\rbrack}
\cr=-4{{\hat{b}}_3}{\lbrace}\oint{{dz}\over{2i\pi}}e^{\phi-\chi}\psi_m
\partial{X^m}(z),
c\partial{c}
{e^\chi}\partial\chi
(\psi_{m_1}...\psi_{m_5}(\psi\partial{X})^{\perp}\cr+
\psi_{{\lbrack}m_1}...\psi_{m_4}\partial{X_{m_5\rbrack}}(\partial\phi
-\partial\chi)
\cr+\psi_{{\lbrack}m_1}...\psi_{m_4}\partial^2{X_{m_5\rbrack}}+
i\psi_{m_1}...\psi_{m_5}((k^{\perp}\psi)(\partial\phi-\partial\chi)+
(k^{\perp}\partial\psi))){e^{ik^{\perp}X}}(w)\rbrace\cr=-4
:{{\hat{b}}_3}\gamma{c}\partial{c}F_{matter}(X,\psi)F_{ghost}(\phi,\chi):(w)\cr
=-4:({{\hat{b}}_3}{c}\partial{c}):(w):{\gamma}F_1(X,\psi,\phi,\chi):(w)
}}
where $F_1(X,\psi,\phi,\chi)$ denotes the expression
that depends entirely on matter fields $X,\psi $ and
bosonized superconformal ghosts
$\phi,\chi$, i.e. it contains no dependence on the fermionic ghost fields.
The precise expression for $F_{1}(X,\psi,\phi,\chi)$
is skipped for the sake of shortness since it is acted on trivially by
${{\hat{b}}_3}$ and plays no role in further calculations.
Now, we have
\eqn\grav{\eqalign{:{{\hat{b}}_3}{c}\partial{c}:(w)=
lim_{u\rightarrow{w}}
\lbrack\oint{{dz}\over{2i\pi}}(z-u)^{4}b(z),:c\partial{c}:(w)\rbrack\cr
=lim_{u\rightarrow{w}}(-4c(w)(u-w)^3+\partial{c(w)}(u-w)^4)=0}}
and therefore
\eqn\grav{\eqalign{{\lbrack}\oint{{dz}\over{2i\pi}}e^{\phi-\chi}
\psi_m\partial{X^m}(z),C_2(w){\rbrack}
=0}}
Finally, let us show that the commutator of $C_2$ with
 the stress-energy part of
$Q_{BRST}=\oint{{dw}\over{2i\pi}}(cT-b:c\partial{c}:)(w)$ also vanishes.
Again, as previously we have
\eqn\grav{\eqalign{\lbrace{{\hat{b}}_3}\oint{{dw}\over{2i\pi}}
(cT-b:c\partial{c}:)(w)\rbrace
=lim_{u\rightarrow{w}}\oint{{dw}\over{2i\pi}}(-12(u-w)^2+O((u-w)^3))=0}}
and therefore, using the fact that
$\lbrack\oint{{dz}\over{2i\pi}}(cT-b:c\partial{c}:)(z)
,c\partial{c}(w)\rbrack=0$
we write
\eqn\grav{\eqalign{\lbrack\oint{{dz}\over{2i\pi}}(cT-b:c\partial{c}:)(z),
C_2(w)\rbrack\cr=
-4{{\hat{b}}_3}\lbrace\oint{{dz}\over{2i\pi}}(cT-b:c\partial{c}:)(z);
c\partial{c}
{e^\chi}\partial\chi
(\psi_{m_1}...\psi_{m_5}(\psi\partial{X})^{\perp}\cr+
\psi_{{\lbrack}m_1}...\psi_{m_4}\partial{X_{m_5\rbrack}}
(\partial\phi-\partial\chi)
\cr+\psi_{{\lbrack}m_1}...\psi_{m_4}\partial^2{X_{m_5\rbrack}}+
i\psi_{m_1}...\psi_{m_5}((k^{\perp}\psi)(\partial\phi-\partial\chi)+
(k^{\perp}\partial\psi))){e^{ik^{\perp}X}}(w)\rbrace\cr
=-4{{\hat{b}}_3}c\partial{c}\partial^2{c}F(X,\psi,\phi,\chi)(w)=
-4:({{\hat{b}}_3}c\partial{c}\partial^2{c}):(w):F(X,\psi,\phi,\chi):(w)\cr=
-4:F(X,\psi,\phi,\chi):(w){\times}lim_{u\rightarrow{w}}(12(u-w)^2{c}\partial{c}
\cr-4(u-w)^3{c}\partial^2{c}+(u-w)^4\partial{c}\partial^2{c})=0}}
and thus
\eqn\grav{\eqalign{\lbrack\oint{{dz}\over{2i\pi}}(cT-b:c\partial{c}:)(z),
C_2(w)\rbrack=0}}
Finally, putting together the pieces (19),(28),(30) and (33) we have

\eqn\grav{\eqalign{\lbrack{Q_{BRST},C_2(w)}{\rbrack}=
-{1\over2}
e^{2\phi-\chi}
(\psi_{m_1}...\psi_{m_5}(\psi\partial{X})^{\perp}+
\psi_{{\lbrack}m_1}...\psi_{m_4}\partial{X_{m_5\rbrack}}(\partial\phi-
\partial\chi)
\cr+\psi_{{\lbrack}m_1}...\psi_{m_4}\partial^2{X_{m_5\rbrack}}+
i\psi_{m_1}...\psi_{m_5}((k^{\perp}\psi)(\partial\phi-\partial\chi)+
(k^{\perp}\partial\psi))){e^{ik^{\perp}X}}(w)\cr\equiv
-\lbrack\oint{{dz}\over{2i\pi}}e^{\phi-\chi}G_{matter}(z),
{e^{\phi}}\psi_{m_1}...\psi_{m_5}e^{ik^{\perp}X}:(w)\rbrack}}
where $G_{matter}(z)={1\over2}\psi_m\partial{X^m}(z)$ is
the worldsheet matter supercurrent.
So we have shown that adding $C_2(w)$ to
${e^{\phi}}\psi_{m_1}...\psi_{m_5}e^{ik^{\perp}X}(w)$
indeed annihilates the first BRST non-invariance, originating from the
 commutator
$\lbrack\oint{{dz}\over{2i\pi}}{\gamma}G_{matter}(z),{e^{\phi}}\psi_{m_1}...
\psi_{m_5}e^{ik^{\perp}X}(w){\rbrack}$.
Now let us find the expression for the  $C_1(w)$ ghost counterterm
which shall compensate for the
non-invariance originating from the commutator of ${e^{\phi}}\psi_{m_1}...
\psi_{m_5}e^{ik^{\perp}X}(w)$
with the ghost supercurrent term of $Q_{BRST}$
We will be looking for the ansatz for $C_1(w)$ in the form:
\eqn\grav{\eqalign{C_1(w)={({\hat{cT_\chi}})_7}
\partial{b}{b}{P^{(1)}_{\alpha\phi-\alpha\chi+\rho\sigma}}e^{2\phi-\chi}
\psi_{m_1}...\psi_{m_5}
{(\psi\partial^2{X})^{\perp}}e^{ik^{\perp}X}(w)}}
with
\eqn\lowen{T_\chi=((\partial\chi)^2+\lambda\partial^2\chi))}
where $\alpha$,$\rho$,$\lambda$
are some real numbers which are to be determined from the condition that
$C_1(w)$ has suitable commutation relations with  $Q_{BRST}$ to cancel the
above mentioned BRST non-invariance;
 the weight 1 polynomial
${P^{(1)}_{\alpha\phi-\alpha\chi+\rho\sigma}}=\alpha\partial\phi-
\alpha\partial\chi+\rho\partial\sigma$
is defined according to (24).
Analogously to (20), the operator ${({\hat{cT_\chi}})_7}$ is defined so
that
\eqn\lowen{
{({\hat{cT_\chi}})_n}A(w)=
lim_{u\rightarrow{w}}
\lbrack\oint{{dz}\over{2i\pi}}(z-u)^{7}cT_\chi(z),A(w)\rbrack}
for any bosonic operator A; if A is fermionic, the commutator is replaced by
anticommutator, as usual.
Let us start with calculating the commutator of $C_1(w)$ with the matter
supercurrent term of
 $Q_{BRST}$.
It is important for our calculation that
\eqn\grav{\eqalign{\gamma(z){{P^{(1)}_{\alpha\phi-\alpha\chi+\rho\sigma}}
(w)\equiv{e^{\phi-\chi}}
(z){P^{(1)}_{\alpha\phi-\alpha\chi+\rho\sigma}}(w)}
\cr\sim{{\alpha-\alpha}\over{z-w}}+{O((z-w)^0)}
 \sim{O((z-w)^0)}}}
i.e. there is no singularity (simple pole) in the O.P.E.
of ${P^{(1)}_{\alpha\phi-\alpha\chi+\rho\sigma}}$
with $\gamma$.
Also using the fact that, as before,
the ${({\hat{cT_\chi}})_7}$-operator commutes with the matter supercurrent
term of $Q_{BRST}$, we  evaluate the commutator and obtain:

\eqn\grav{\eqalign{{\lbrack}\oint{{dz}\over{2i\pi}}
e^{\phi-\chi}\psi_m\partial{X^m}(z),C_1(w){\rbrack}
\cr=
{({\hat{cT_\chi}})_7}\partial{b}{b}{P^{(1)}_{\alpha\phi-\alpha\chi+\rho\sigma}}
e^{3\phi-2\chi}\psi_{m_1}...\psi_{m_5}e^{ik^{\perp}X}(w)
\cr\times\lbrack{10P^{(4)}_{\phi-\chi}}
+{P^{(2)}_{\phi-\chi}}\lbrack(ik\partial^2{X})^{\perp}
+{(\partial{\psi}\psi)^{\perp}}\rbrack\cr+
{P^{(1)}_{\phi-\chi}}
{\lbrack}(\partial{X}\partial^2{X})^{\perp}+
(\partial^2\psi\psi)^{\perp}\rbrack\cr
+{1\over2}(\partial^2{\psi}\partial\psi)^{\perp}+
{1\over3}(\partial^3{\psi}\psi)^{\perp}+
i{(k\partial^2{X})^{\perp}}{(\partial\psi\psi)^{\perp}}\rbrack\cr
+{({\hat{cT_\chi}})_7}
\partial{b}{b}{P^{(1)}_{\alpha\phi-\alpha\chi+\rho\sigma}}e^{3\phi-2\chi}
\psi_{{\lbrack}m_1}...\psi_{m_4}\partial{X_{m_5\rbrack}}
(\partial\phi-\partial\chi)
\cr+\psi_{{\lbrack}m_1}...\psi_{m_4}
\partial^2{X_{m_5\rbrack}}{e^{ik^{\perp}X}}(w)
\equiv{({\hat{cT_\chi}})_7}{L_1(w)}}}
Next, we need to evaluate the commutator
${({\hat{cT_\chi}})_7}{L_1(w)}$.
Let us make several observations about (32) which simplify things
significantly.
The operator product of  $cT_\chi(z)$ and $L_1(w)$ contains
singular terms with maximum singularity order of  $(z-w)^{-9}$
But because of the factor of  $(z-u)^{7}$ in the definition (37) of
${({\hat{cT_\chi}})_7}$
it is clear that the only  non-vanishing contributions
to this commutator will come from the
 the O.P.E. terms
of  $cT_\chi(z)$ and $L_1(w)$ having
 the  singularity order of $at$  $least$  $(z-w)^{-8}$;
any contributions originating
from less singular terms of this O.P.E.will vanish after
evaluating the contour integral, since they will contain
factors of the type ${lim}_{u\rightarrow{w}}(u-w)^{8-n}=0$
where $n$ is the order of the O.P.E. singularity.
Therefore it is clear that out of all the terms
that constitute the long expression (39) the  only  term
 contributing to the commutator
${({\hat{cT_\chi}})_7}{L_1(w)}$ is the one that contains the weight
4 ghost polynomial
$P^{(4)}_{\phi-\chi}$
 where, in accordance with the definition (24)
\eqn\grav{\eqalign{{P^{(4)}_{\phi-\chi}}=
{1\over{24}}(\partial^4\phi-\partial^4\chi)
+{1\over{6}}(\partial^3\phi-\partial^3\chi)(\partial\phi-\partial\chi)\cr
+{1\over{8}}(\partial^2\phi-\partial^2\chi)^2
+{1\over{4}}(\partial^2\phi-\partial^2\chi)(\partial\phi-\partial\chi)^2
+{1\over{24}}(\partial\phi-\partial\chi)^4}}
Fortunately for us there are no
terms in (39) with the weight 3 $P^{(3)}_{\phi-\chi}$ polynomial factor;
if such  terms were present , they would also have contributed
to ${({\hat{cT_\chi}})_7}{L_1(w)}$,
spoiling the entire construction. As for the terms with the
$P^{(2)}_{\phi-\chi}$
factor, they all have  maximum singularity order of  $(z-w)^{-7}$
in the O.P.E. with $cT_\chi(z)$;  terms with the  $P^{(1)}_{\phi-\chi}$
give the  maximum singularity order of  $(z-w)^{-6}$, i.e.
as has been explained above, these
terms are irrelevant, as they do not  contribute to the commutator.
Having explained this all,
 let us finally evaluate the commutator of ${({\hat{cT_\chi}})_7}$
  with $L_1(w)$,
using (37) and (38), as well as the O.P.E's:
\eqn\grav{\eqalign{ e^\sigma(z){P^{(1)}_{\alpha\phi-\alpha\chi+\rho\sigma}}(w)
\sim{-}{{\rho}\over{z-w}}
e^\sigma(w)+:({P^{(1)}_{\alpha\phi-\alpha\chi+\rho\sigma}}
-\rho\partial\sigma)e^\sigma:(w)+...
\cr={-}{{\rho}\over{z-w}} e^\sigma(w)+:{P^{(1)}_{\alpha\phi-\alpha\chi}}
e^\sigma:(w)+...\cr
\partial\chi(z)\partial^n\chi(w)\sim{{n!}\over{(z-w)^{n+1}}}+
:\partial\chi\partial^n\chi:(w)+...\cr
\partial\chi(z)e^{2\phi-\chi}(w)\sim{-}{{1}\over{z-w}}+
:\partial{\chi}e^{2\phi-\chi}:(w)+...\cr
etc.}}
We have:
\eqn\grav{\eqalign{
{lim}_{u\rightarrow{w}}
10\oint{{dz}\over{2i\pi}}
(z-u)^{7}:c((\partial\chi)^2+\lambda\partial^2\chi):(z)
{P^{(4)}_{\phi-\chi}}\partial{b}{b}{P^{(1)}_{\alpha\phi-\alpha\chi+\rho\sigma}}
\cr{\times}e^{3\phi-2\chi}\psi_{m_1}...\psi_{m_5}
e^{ik^{\perp}X}(w)\cr=
{{lim}_{u\rightarrow{w}}}
10\oint{{dz}\over{2i\pi}}
{{(z-u)^{7}}\over{(z-w)^8}}
be^{3\phi-2\chi}\psi_{m_1}...\psi_{m_5}e^{ik^{\perp}X}(w)\cr
\lbrace{3}({P^{(1)}_{\alpha\phi-\alpha\chi+\rho\sigma}}-\rho\partial\sigma)
-\rho(2\partial\phi-3\partial\chi)+2\alpha\partial\sigma-2\alpha(\partial\phi
-\partial\chi)\rbrace\cr=10{P^{(1)}_{(\alpha-2\rho)\phi+
(3\rho-\alpha)\chi+2\alpha\sigma}}
be^{3\phi-2\chi}\psi_{m_1}...\psi_{m_5}e^{ik^{\perp}X}(w)}}
Next,
\eqn\grav{\eqalign{
{lim}_{u\rightarrow{w}}
10\oint{{dz}\over{2i\pi}}
(z-u)^{7}:c\partial^2\chi:(z)
P^{(4)}_{\phi-\chi}\partial{b}{b}{P^{(1)}_{\alpha\phi-\alpha\chi+\rho\sigma}}
\cr{\times}e^{3\phi-2\chi}\psi_{m_1}...\psi_{m_5}
e^{ik^{\perp}X}(w)\cr=
10{P^{(1)}_{(5\alpha-4\rho)\phi-(4\rho+5\alpha)\chi}}
be^{3\phi-2\chi}\psi_{m_1}...\psi_{m_5}e^{ik^{\perp}X}(w)}}
Adding these two contributions together gives:
\eqn\grav{\eqalign{
{\lbrack}\oint{{dz}\over{2i\pi}}e^{\phi-\chi}\psi_m\partial{X^m}(z);
C_1(w){\rbrack}\equiv
({\hat{cT_\chi}})_7{L_1(w)}\cr=
{lim}_{u\rightarrow{w}}
10\oint{{dz}\over{2i\pi}}
(z-u)^{7}:c(\partial\chi\partial\chi+\lambda\partial^2\chi):(z)
{P^{(4)}_{\phi-\chi}}\partial{b}{b}{P^{(1)}_{\alpha\phi-\alpha\chi+\rho\sigma}}
\cr{\times}e^{3\phi-2\chi}\psi_{m_1}...\psi_{m_5}
e^{ik^{\perp}X}(w)\cr=
10{P^{(1)}_{((\alpha-2\rho)+\lambda(5\alpha-4\rho))\phi+
((3\rho-\alpha)-\lambda(4\rho+5\alpha))\chi+2\alpha\sigma}}
be^{3\phi-2\chi}\psi_{m_1}...\psi_{m_5}e^{ik^{\perp}X}(w)}}
Finally, we have to choose the coefficients $\alpha$, $\rho$ and $\lambda$
so that $({\hat{cT_\chi}})_7{L_1(w)}$ could annihilate the second
BRST non-invariance (21) of the vertex
$e^{\phi}\psi_{m_1}...\psi_{m_5}e^{ik^{\perp}X}(w)$,
originating from its commutation with the ghost supercurrent term of the
 $Q_{BRST}$.
The obvious condition for that is given by
\eqn\grav{\eqalign{
{P^{(1)}_{((\alpha-2\rho)+\lambda(5\alpha-4\rho))\phi+((3\rho-\alpha)
-\lambda(4\rho+5\alpha))\chi+2\alpha\sigma}}
={P^{(1)}_{2\phi-2\chi-\sigma}}(w)}}
from which we immediately have
\eqn\lowen{\alpha=-{1\over2}}
and the following system of two equations for  $\rho$ and $\lambda$:
\eqn\grav{\eqalign{-{1\over2}-2\rho-{5\over2}\lambda-4\lambda\rho=2\cr
{1\over2}+3\rho+{5\over2}\lambda-4\lambda\rho=-2}}
which has the following solutions:
\eqn\grav{\eqalign{\rho=0,\lambda=-1\cr
\rho={7\over8},\lambda=-{{15}\over8}}}
These two solutions correspond to physically identical
gauge choices for $C_1$. It is natural to choose the
gauge first solution so that
\eqn\lowen{T_\chi=\partial\chi\partial\chi-\partial^2\chi}
is just twice the stress-energy tensor for the $\chi$-field;
and the expression for
$C_1$ is obtained to be:
\eqn\grav{\eqalign{C_1(w)={({\hat{cT_\chi}})_7}
{P^{(1)}_{-{1\over2}\phi+{1\over2}\chi}}\partial{b}be^{2\phi-\chi}
\psi_{m_1}...\psi_{m_5}
(\psi\partial^2{X})^{\perp}e^{ik^{\perp}X}}}
Because of the factor of $10$ in (32) this operator should of course
be multiplied by the appropriate numerical factor (given by $-{1\over{20}}$)
 to compensate the BRST non-invariance.
To summarize (15), (44) and (50), we have shown that
\eqn\grav{\eqalign{-{1\over{20}}
{\lbrack}\oint{{dz}\over{2i\pi}}e^{\phi-\chi}\psi_m\partial{X^m}(z),
C_1(w){\rbrack}
=
-{1\over4}{\lbrack}\oint{{dz}\over{2i\pi}}e^{2\phi-2\chi}b(z);
e^{\phi}\psi_{m_1}...\psi_{m_5}
e^{ik^{\perp}X}(w)\lbrack}}
Finally, to prove that $-{1\over{20}}C_1(w)$ is indeed the operator
 to insure (along with $C_2(w)$) the BRST-invariance of the picture $+1$
five-form, we still need to show that $C_1(w)$ commutes with all other terms in
$Q_{BRST}$.
Let us start with the commutator
${\lbrack}\oint{{dz}\over{2i\pi}}e^{2\phi-2\chi}b(z),C_1(w)\rbrack$.
This commutator vanishes since, as usual,
\eqn\lowen{\lbrace\oint{{dz}\over{2i\pi}}e^{2\phi-2\chi}b(z),
{({\hat{cT_\chi}})_7}\rbrace=0}
and
\eqn\grav{\eqalign{b(z):\partial{b}b:(w)\sim{(z-w)^2}e^{-3\sigma}(w)+
O((z-w)^3)\cr
:e^{2\phi-2\chi}:(z):e^{2\phi-2\chi}:(w)\sim{(z-w)^{-2}}e^{4\phi-3\chi}
+O((z-w)^{-1})\cr
e^{2\phi-2\chi}{P^(1)_{-{1\over2}\phi+{1\over2}\chi}}(w)\sim{O((z-w)^{0})}}}
Therefore
\eqn\grav{\eqalign{{\lbrace}\oint{{dz}\over{2i\pi}}e^{2\phi-2\chi}b(z),
{P^(1)_{-{1\over2}\phi+{1\over2}\chi}}\partial{b}be^{2\phi-\chi}
\psi_{m_1}...\psi_{m_5}e^{ik^{\perp}X}\rbrace
=0}}
and hence
\eqn\grav{\eqalign{{\lbrack}\oint{{dz}\over{2i\pi}}e^{2\phi-2\chi}b(z);
C_1(w)\rbrack=0}}
Next, let us focuse on the commutator of
$C_1(w)$ with $\oint{{dz}\over{2i\pi}}(cT_{ghost}-b:c\partial{c})$
of $Q_{BRST}$.
Again,
\eqn\lowen{\lbrace\oint{{dz}\over{2i\pi}}(cT_{ghost}-b:c\partial{c}:)(z)
{({\hat{cT_\chi}})_7}(w)\rbrace=0}
since
\eqn\lowen{(cT_{ghost}-b:c\partial{c}:)(z){({\hat{cT_\chi}})_7}(w)
\sim:c\partial{c}:(w)
\times{lim}_{u\rightarrow{w}}(u-w)^4+O((u-w)^5)=0}
Next, it is easy to check that
\eqn\grav{\eqalign{\lbrace
\oint{{dz}\over{2i\pi}}(cT_{ghost}-b:c\partial{c})(z),
{P^(1)_{-{1\over2}\phi+{1\over2}\chi}}\partial{b}be^{2\phi-\chi}
\psi_{m_1}...\psi_{m_5}
(\psi\partial^2{X})^{\perp}e^{ik^{\perp}X}(w)\rbrace\cr
={G^{(4)}_{\phi,\chi,\sigma}}be^{2\phi-\chi}\psi_{m_1}...\psi_{m_5}
(\psi\partial^2{X})^{\perp}e^{ik^{\perp}X}(w)}}
where ${G^{(4)}_{\phi,\chi,\sigma}}$ is the conformal weight 4 polynomial
consisting of  derivatives of $\phi$, $\chi$ and $\sigma$,
which exact form is not important and is not given here
for the sake of brevity.
It is easy to see from the simple dimensiaonal analysis that
the operator product between $cT_\chi:(z)$  and
${G^{(4)}_{\phi,\chi,\sigma}}(w)$ must be given by
\eqn\grav{\eqalign{cT_\chi(z)
{G^{(4)}_{\phi,\chi,\sigma}}(w)\sim
{a\over{(z-w)^6}}c(w)+O((z-w)^3)}}
where $a$ is some number, i.e. the most singular term in this O.P.E
is of order of
${(z-w)^{-6}}$
Therefore
\eqn\grav{\eqalign{{({\hat{cT_\chi}})_7}{G^{(4)}_{\phi,\chi,\sigma}}
be^{2\phi-\chi}\psi_{m_1}...\psi_{m_5}
(\psi\partial^2{X})^{\perp}e^{ik^{\perp}X}(w)\cr=
{lim}_{u\rightarrow{w}}\oint{{dz}\over{2i\pi}}(z-u)^7{cT_\chi(z)}
{G^{(4)}_{\phi,\chi,\sigma}}be^{2\phi-\chi}\psi_{m_1}...\psi_{m_5}
(\psi\partial^2{X})^{\perp}e^{ik^{\perp}X}(w)
\cr=e^{2\phi-\chi}\psi_{m_1}...\psi_{m_5}
(\psi\partial^2{X})^{\perp}e^{ik^{\perp}X}(w)
{\times}{lim}_{u\rightarrow{w}}\oint{{dz}\over{2i\pi}}(z-u)^7(z-w)^{-7}
\cr=7e^{2\phi-\chi}\psi_{m_1}...\psi_{m_5}
(\psi\partial^2{X})^{\perp}e^{ik^{\perp}X}(w)
{\times}{lim}_{u\rightarrow{w}}(u-w)=0}}
so we have just established
\eqn\lowen{\lbrack\oint{{dz}\over{2i\pi}}(cT_{ghost}-b:c\partial{c}:)(z);
C_1(w)\rbrack=0}.
To conclude our derivation, we finally have to show that $C_1$ commutes
 with the last remaining piece of
$Q_{BRST}$ given by $\oint{{dz}\over{2i\pi}}cT_{matter}(z)$.
Again, as previously,
\eqn\lowen{\lbrace\oint{{dz}\over{2i\pi}}cT_{matter}(z)
{({\hat{cT_\chi}})_7}(w)\rbrace=0}
Next,
\eqn\grav{\eqalign{\lbrace\oint{{dz}\over{2i\pi}}cT_{matter}(z),
{P^{(1)}_{-{1\over2}\phi+{1\over2}\chi}}\partial{b}
be^{2\phi-\chi}\psi_{m_1}...\psi_{m_5}
(\psi\partial^2{X})^{\perp}e^{ik^{\perp}X}(w)\rbrace\cr=
be^{2\phi-\chi}\psi_{m_1}...\psi_{m_5}
(\psi\partial^2{X})^{\perp}e^{ik^{\perp}X}(w)\cr
\times(2i{(k\psi)^{\perp}}P_{\sigma}^{(5)}+
2{(\partial{X}\psi)^{\perp}}P_\sigma^{(4)}
+5{(\partial^2{X}\psi)^{\perp}}P_\sigma^{(3)}
+{(\partial^3{X}\psi)^{\perp}}P_\sigma^{(2)}
\cr+{1\over3}{(\partial^4{X}\psi)^{\perp}}P_\sigma^{(1)}
+{1\over{12}}{(\partial^5{X}\psi)^{\perp}}
\cr+
i{(\partial^2{X}\psi)^{\perp}}{(k\partial{X})^{\perp}}
P_\sigma^{(2)}+i{(\partial^2{X}\psi)^{\perp}}{(k\partial^2{X})^{\perp}}
P_\sigma^{(1)}
+{i\over2}{(\partial^2{X}\psi)^{\perp}}{(k\partial^3{X})^{\perp}}\cr
+{(\partial^2{X}\partial\psi)^{\perp}}P_\sigma^{(2)}+
{3\over4}{(\partial^2{X}\partial^2\psi)^{\perp}}P_\sigma^{(1)}
\cr+{1\over3}{(\partial^2{X}\partial^3\psi)^{\perp}}
+{(\partial^2{X}\psi)^{\perp}}{(\partial{\psi}\psi)^{\perp}}P_\sigma^{(1)}
+{(\partial^2{X}\psi)^{\perp}}\partial{(\partial{\psi}\psi)^{\perp}}\rbrace\cr+
{P^{(1)}_{-{1\over2}\phi+{1\over2}\chi}}\partial{b}be^{2\phi-\chi}
(\psi\partial^2{X})^{\perp}\cr\times
{\lbrace}{5\over2}\psi_{m_1}...\psi_{m_5}P_{\sigma}^{(3)}+
\partial(\psi_{m_1}...\psi_{m_5})P_{\sigma}^{(2)}+
{3\over4}\partial^2\psi_{{\lbrack}m_1}\psi_{m_2}...\psi_{m_5\rbrack}
P_\sigma^{(1)}
\cr+{1\over3}\partial^3
\psi_{{\lbrack}m_1}\psi_{m_2}...\psi_{m_5{\rbrack}}\rbrace\cr\equiv{L_2}(w)}}
Now we have to show that
${({\hat{cT_\chi}})_7}{L_2}(w)=0$.
The crucial point to observe is that the O.P.E of the polynomials
$P_\sigma^{(n)}(w)$
with the field $c(z)$ cannot contain terms more singular than
$(z-w)^{-1}$ for any $n$,
even though the polynomials have conformal weight equal to n.
Indeed, since $c(z)c(w)\sim(z-w):c\partial{c}(w)+O((z-w)^2)$, i.e.
the lowest order term in this O.P.E. is of order of
$(z-w)$,
it is clear that the product $\partial_z^nc(z)c(w)\sim{0((z-w)^0)}$,
i.e. non-singular.
But by definition $\partial_z^nc(z){\equiv}n!P_\sigma^{(n)}c(z)$ and
therefore there can be no singularities
other that a simple pole in the O.P.E of
$P_\sigma^{(n)}(w)$ and  $c(z)$ Then, since the O.P.E of $T_\chi$ with
$L_2(w)$ has the most singular part of order
of  $(z-w)^{-3}$ and $c(z)b(w)\sim{(z-w)^{-1}}$ we have
\eqn\grav{\eqalign{cT_\chi(z)L_2(w)\sim{1\over{(z-w)^5}}}}
\eqn\grav{\eqalign{{({\hat{cT_\chi}})_7}{L_2}(w)
\sim{lim}_{u\rightarrow{w}}
\oint{{dz}\over{2i\pi}}(z-u)^7((z-w)^{-5}
+O((z-w)^4)+...)\cr
\sim{lim}_{u\rightarrow{w}}((u-w)^4+O((u-w)^5))=0}}
and
therefore
\eqn\lowen{\lbrace\oint{{dz}\over{2i\pi}}cT_{matter}(z),C_1(w)\rbrace=0}

Summing up (50),(51),(55),(66)
we obtain
\eqn\grav{\eqalign{-{1\over{20}}{\lbrack}Q_{BRST},C_1(w){\rbrack}\equiv
=-{1\over4}{\lbrack}\oint{{dz}\over{2i\pi}}e^{2\phi-2\chi}b(z),
e^{\phi}\psi_{m_1}...\psi_{m_5}e^{ik^{\perp}X}(w)}}
We have shown that the operator given by $-{1\over{20}}C_1(w)$
indeed compensates for the second BRST non-invariance of
$e^{\phi}\psi_{m_1}...\psi_{m_5}e^{ik^{\perp}X}(w)$ and
using (19),(34),(50) and (67) we see that the total combination
$e^{\phi}\psi_{m_1}...\psi_{m_5}
e^{ik^{\perp}X}(w)-2C_2(w)-{1\over{20}}C_1(w)$ is BRST-invariant.
To summarize our calculation we have shown that the BRST-invariant
expression for the 5-form
operator  at the $+1$-picture is given by:
\eqn\grav{\eqalign{V_5^{(+1)}(w)=\oint{{dw}\over{2i\pi}}\lbrace
e^{\phi}\psi_{m_1}...\psi_{m_5}e^{ik^{\perp}X}(w)
\cr-2{\hat{b_3}}c\partial{c}{e^\chi}\partial\chi
(\psi_{m_1}...\psi_{m_5}(\psi\partial{X})^{\perp}\cr+
\psi_{{\lbrack}m_1}...\psi_{m_4}\partial{X_{m_5\rbrack}}(\partial\phi
-\partial\chi)
\cr+\psi_{{\lbrack}m_1}...\psi_{m_4}\partial^2{X_{m_5\rbrack}}+
i\psi_{m_1}...\psi_{m_5}((k^{\perp}\psi)(\partial\phi-\partial\chi)+
(k^{\perp}\partial\psi))){e^{ik^{\perp}X}}(w)\cr-{1\over{20}}
{({\hat{cT_\chi}})_7}
{P^(1)_{-{1\over2}\phi+{1\over2}\chi}}\partial{b}be^{2\phi-\chi}
\psi_{m_1}...\psi_{m_5}
(\psi\partial^2{X})^{\perp}e^{ik^{\perp}X}\rbrace}}
This concludes the derivation of the BRST-invariant 5-form vertex operator
at the $+1$-picture as well as our discussion of the BRST-invariance
 of the brane-like states.

\centerline{\bf 2. BRST non-triviality of the Brane-like states}

The specific property of the picture $-3$ five-form vertex operator
in the unintegrated form, as well
as of the two-form (2) is that their picture changing transformation vanishes
(for the two-form this is true only in he zero momentum case).
This may cause one to suspect that these two states are actually BRST-trivial.
Indeed, there exists a conventional wisdom, implying that if
a BRST-invariant operator $V$ satisfies $:\Gamma{V}:=0$ then $V$
is BRST-trivial,
i.e. there exists an operator A such that
$V=\lbrace{Q_{BRST}}, V\rbrace$. Below we will show that this
 conventional wisdom is wrong and
comment on what actually goes wrong in the standard proof of this claim.
First of all, that this conventional wisdom is wrong is easy to see
from the following simple example.
Consider an open string three-point
function $<V_F^{-3/2}(k)V_F^{-1/2}(-k)V_B(0)>$
where
\eqn\grav{\eqalign{V_F^{-3/2}(k)=u_\alpha(k)ce^{-{3\over2}\phi}
\Sigma_\alpha{e^{ikX}}(z_1)\cr
V_F^{-1/2}(-k)=v_\alpha(-k)ce^{-{1\over2}\phi}\Sigma_\alpha{e^{-ikX}}(z_1)\cr
V_B(0)=e_m(0)c\partial{X^m}}}
Here $V_F^{-1/2}(-k)$ is the standard Ramond vertex at the
${-1\over2}$-picture with
the usual on-shell condition $k\gamma^m_{\alpha\beta}u^\beta(k)=0$,
where $\gamma^m$ are symmetric $16\times{16}$  ten-dimensional gamma-matrices.
$V_B(0)$ is a photon vertex operator which for simplicity we
take at zero momentum
(therefore there are no transversality conditions on the $e_m(0)$
polarization vector);
 and ${V_F^{-3/2}}(k)$ is the Ramond vertex operator at the picture $-3/2$
which
polarization spinor is choosen to impose the condition
$k\gamma^m_{\alpha\beta}u^\beta(k)=0$, even though in principle
such a condition is not necessary
in case of the picture $-3/2$ Ramond operator (unlike the picture $-1/2$ case).
If, however, we do choose such a condition,
it is easy to see that the picture-changing transformation of
$V_F^{-3/2}(k)$ vanishes:
\eqn\lowen{:{\Gamma}V_F^{-3/2}(k):=i(k_m\gamma^m_{\alpha\beta}u_\alpha)
ce^{-{1\over2}\phi}\Sigma_\beta{e^{ikX}}=0}
due to the condition that we have imposed on the polarization spinor $u(k)$
However, it is elementary to see that
the three-point correlation function, described above,
does not vanish but is equal to:
\eqn\lowen{<V_F^{-3/2}(k)V_F^{-1/2}(-k)V_B(0)>=i(ke(0))(u(k)v(-k))}
But since all the operators:
$V_F^{-3/2}(k)$, $V_F^{-1/2}(-k)$ and $V_B(0)$
are BRST-invariant and their three-point function is nonzero, this means that
the operator $V_F^{-3/2}(k)$ cannot be BRST-trivial, even though its
picture-changing transform vanishes.
Having illustrated the clear error in the
conventional wisdom on this simple example,
let us discuss in more details  the origin of this error.
The standard proof of the claim,  on which the mistaken wisdom is based,
is the following.
One takes a  BRST-invariant operator $V$, such that
\eqn\lowen{:\Gamma{V}:=0}
and multiplies this identity by an inverse picture-changing operator
\eqn\lowen{{\Gamma}^{-1}={2\over3}ce^{\chi-2\phi}\partial\chi}.
Then, by claiming
\eqn\lowen{:{\Gamma}^{-1}\Gamma{V}:=0}
and
\eqn\lowen{:{\Gamma}^{-1}\Gamma:=\lbrace{Q_{BRST}},A\rbrace}
where A is some operator,
and using the BRST-invariance of V,
one writes
\eqn\lowen{0=:{\Gamma}^{-1}\Gamma{V}:=V+\lbrace{Q_{BRST}},A\rbrace{V}
=V+\lbrace{Q_{BRST}},AV\rbrace}
and hence
claiming the BRST triviality of any operator V with vanishing
picture-changing transform:
\eqn\lowen{V=-\lbrace{Q_{BRST}},AV\rbrace}
It is easy to see, however, that such a proof, although standard, contains
consistency problems and contradictions, which make it impossible
to apply it to the  case of the picture $-3$ five-form.
First of all, it must be remembered that the identity
(69) is true only in the normally ordered sense; strictly speaking,
there are also higher order terms in the full O.P.E
and in general one is not allowed to drop these terms
in the product of three operators ${\Gamma}^{-1}$,
${\Gamma}$ and $V$, substituting the product ${\Gamma}^{-1}\Gamma{V}$
with its normally ordered part.
Such a substitution, neglecting the higher order terms
in the O.P.E of ${\Gamma}$ and $V$,
 is particularly
incorrect if
${\Gamma}^{-1}$ has a singular O.P.E. with V - and
this is exactly what happens in case of
${\Gamma}^{-1}$ and the picture $-3$ five-form (2).

 More generally, substituting
$AB$ with $:AB:$ in the product of three operators A, B and C
is not correct. Moreover, if these operators are such that
$:BC:=0$, this would not necessarily imply $:ABC:=0$
To illustrate the above remarks, consider a simple example
$A=e^\phi{b}$,$B=e^{-\phi}c$, $C=e^{n\phi}$ where
n is some positive number.
Clearly,
\eqn\lowen{:AB:\sim{1}}
 but at the same time
\eqn\lowen{:BC:\sim{0}}
Now if one follows the the logic of the standard proof of (77),
neglecting
the higher order terms,
one can multiply the identity (79) by A
to obtain
\eqn\lowen{ABC\sim{0}}
At the same time, using (78) one would get
\eqn\lowen{:ABC:\sim{C}}
and hence
\eqn\lowen{{C}\sim{0}}
which clearly is a contradiction.
As was noted above, the reason for this contradiction is
that when one is computing the normally ordered
expression for ABC, one cannot substitute AB with its normally ordered
part if the O.P.E. between A and C (or between B and C)
is singular . Since this is clearly fulfilled in case of the inverse
picture-changing operator and the  picture $-3$ five-form (2),
the vanishing picture-changing transformation
of this  five-form does NOT imply its BRST triviality.

BRST non-triviality condition, however, does impose strong
constraints on the propagation of the 5-form,
constraining it to propagate in lower dimensional subspace
of the $R^{10}$
Below we shall give a detailed analysis of these constraints.
To start with, there are two and only two possible sources of potential
BRST triviality threat for the picture -3 five-form (2). Namely,
this threat is
coming from commutations of dimension 1 primary fields $U_1$ and
$U_2$,
such that commutators of $U_1$ with matter supercurrent part of $Q_{BRST}$
and those of $U_2$ with the ghost supercurrent part of the BRST charge.
The only possible expression for $U_1$ is given by:
\eqn\grav{\eqalign{U_1(w)=\oint{{dw}\over{2i\pi}}
e^{\chi-4\phi}\partial\chi\psi_{m_1}...\psi_{m_5}(\psi\partial{X})
e^{ik^{\perp}X}(w)}}
Indeed, as is easy to check
\eqn\grav{\eqalign{\lbrack\oint{{dw}\over{2i\pi}}\gamma(\psi\partial{X});
U_1(w)\rbrack=e^{-3\phi}\partial\chi\psi_{m_1}...\psi_{m_5}
e^{ik^{\perp}X}(w)}}
and
\eqn\grav{\eqalign{\lbrack\oint{{dw}\over{2i\pi}}\gamma^2b,U_1(w)\rbrack
=0}}
Therefore in order to insure the BRST non-triviality
of the picture $-3$ five-form operator
one must insure that $U_1$ does not commute with the
stress-energy part of $Q_{BRST}$. This in turn imposes constraints
on the propagation of the five-form, restricting
$k^{\perp}$ to propagate in five-dibensional subspace,
orthogonal to the one spanned by the indices
${m_1}$...${m_5}$. Indeed,  the supercurrent factor
$(\psi\partial{X})$ in the expression for the  $U_1$-operator
clearly directed orthogonally to the subspace spanned by
${m_1}$...${m_5}$, since $\psi_{m_i}^2=0$.
Therefore it is easy to see that
\eqn\grav{\eqalign{\lbrack\oint{{dw}\over{2i\pi}}
(c(T_{matter}+T_{ghost})-b:c\partial{c})(z),
U_1(w)=e^{\chi-4\phi}\partial\chi\psi_{m_1}...\psi_{m_5}e^{ik^{\perp}X}(w)
{\times}(\psi^tk^{\perp}_t)}}
where $t$ denotes the directions orthogonal to
${m_1}$...${m_5}$. Therefore the last commutator
is non-vanishing only if the scalar product $(\psi^tk^{\perp}_t)$ is
non-zero, i.e. $k^{\perp}$ is also directed orthogonally
to the ${m_1}$...${m_5}$ directions.
Equivalently, one may say that if  $k^{\perp}$
is orthogonal to ${m_1}$...${m_5}$, then the  $U_1$-operator
is not a primary field and therefore does not commute with
$\oint{{dw}\over{2i\pi}}(c(T_{matter}+T_{ghost})-b:c\partial{c})(z)$
of  $Q_{BRST}$. We conclude that
BRST non-triviality condition constrains the
propagation of the five-form vertex operator to
the five-dimensional subspace of  $R^{10}$.
Therefore one may think of the five-form as of a multidimensional analogue
of the discrete states that are well-known to occur in two-dimensional
quantum gravity.
Finally, the second threat of the BRST triviality
comes from the operator
\eqn\grav{\eqalign{U_2=(3c{c}\prime\prime\prime\prime
\eta\prime\eta\prime\prime\prime
-2c{c}\prime\prime\prime\eta\prime\eta\prime\prime\prime){e^{-5\phi}}
\psi_{m_1}...\psi_{m_4}e^{ikX}}}
(where we denoted $\eta=e^\chi$)
since, as one can straightforwardly check this operator
is a primary field commuting with
$\oint{{dw}\over{2i\pi}}(c(T_{matter}+T_{ghost})-b:c\partial{c})(z)$
of  $Q_{BRST}$.
It also commutes with
the matter supercurrent part
$\oint{{dw}\over{2i\pi}}\gamma(\psi\partial{X})$
of $Q_{BRST}$, while a simple analysis of conformal dimensions and
of the ghost numbers easily shows that
the ghost supercurrent part of $Q_{BRST}$, given by
$\oint{{dw}\over{2i\pi}}\gamma^2b$ has a simple pole
in the O.P.E. with the integrand of  the $U_2$-operator
and , up to some numerical coefficient, produces exactly
the five-form vertex operator,
which would imply its BRST-triviality.
However, the crucial point is that
this numerical coefficient turns out to be equal to zero,
and therefore, as
there are clearly no other
triviality threats, the five-form is NOT BRST exact
and IS physical. Let us give a detailed proof that indeed this
numerical coefficient is zero, which means
that  $Q_{BRST}$ simply commutes with  $U_2$ and therefore
the  $U_2$ poses no  triviality threat to the five-form.
The proof follows below.
Using the standard bosonization formulae:
\eqn\grav{\eqalign{c={e^\sigma}\cr
b=e^{-\sigma}\cr \eta={e^\chi}\cr \gamma=e^{\phi-\chi}}} Then, as it is
straightforward to check, the normal ordered bosonization formulae for the
two terms constituting the $U_2$-operator
(without the $e^{-5\phi}$-factor
and
the matter part) are given by:
\eqn\grav{\eqalign{{W_1}=3c{c}\prime\prime\prime\prime
\eta\prime\eta\prime\prime
=3{\lbrace}(-{1\over2}(e^{2\sigma})\prime\prime\prime
-3(\sigma\prime\prime{e^{2\sigma}})\prime)(\chi\prime\prime-
\chi\prime\chi\prime)e^{2\chi}\rbrace\cr
{W_2}=2c{c}\prime\prime\prime\eta\prime\eta\prime\prime\prime
=2\lbrace({-}{3\over4}(e^{2\sigma})\prime\prime-{3\over2}
(\sigma)\prime\prime{e^{2\sigma}})(-{1\over4}(e^{2\chi})\prime\prime\prime
+{3\over2}(\chi\prime\prime{e^{2\chi}})\prime)\rbrace}}
where $U_2{\equiv}e^{-5\phi}(W_1-W_2)\psi_{m_1}...\psi_{m_5}e^{ikX}$.

Now we need to evaluate the most singular terms
(of order of $(z-w)^{-11}$) of the O.P.E
of $W_1$ and $W_2$ with $\gamma^2b$ of $Q_{brst}$ because these terms
are
the only ones contributing to the simple pole
(as $e^{2\phi}(z)e^{-5\phi}(w)\sim(z-w)^{10}{e^{-3\phi}}(w)$)
Using the O.P.E's
\eqn\grav{\eqalign{\sigma\prime(z)e^{\alpha\sigma}(w)\sim
{\alpha\over{z-w}}e^{\alpha\sigma}(w)+...\cr
\sigma\prime\prime(z)e^{\alpha\sigma}(w)\sim{-{{\alpha}\over{(z-w)^2}}}
e^{\alpha\sigma}(w)+...\cr
e^{\alpha\sigma}(z)e^{\beta\sigma}(w)\sim(z-w)^{\alpha\beta}
e^{(\alpha+\beta)\sigma}(w)
etc}}
and similarly for the exponents of the type $e^{\alpha\chi}$
we obtain:
\eqn\grav{\eqalign{e^{-2\chi}(z)
(\chi\prime\prime-
\chi\prime\chi\prime)e^{2\chi}(w)\sim
(z-w)^{-4}({2\over{(z-w)^2}}-{4\over{(z-w)^2}})+...\cr
=-2(z-w)^{-6}+...\cr
e^{-\sigma}(z):(-{1\over2}(e^{2\sigma})\prime\prime\prime
-3(\sigma\prime\prime{e^{2\sigma}})\prime)(w):\sim
e^{\sigma}(w)\lbrace{-}{1\over2}\partial_w^3({1\over{(z-w)^2}})
-3\partial_w({1\over{(z-w)^4}}) \cr =-24(z-w)^{-5}}}
and therefore
\eqn\lowen{:\gamma^2b:(z){W_1}(w)\sim
3\times(-2)\times(-24)(z-w)^{-5-6}ce^{2\phi}(w)+...=144(z-w)^{-11}
ce^{2\phi}(w)+...}

Next, consider the  O.P.E. of $\gamma^2b$ with $W_2$.
Again, proceeding as above we have:
\eqn\grav{\eqalign{
e^{-\sigma}(z):({-}{3\over4}(e^{2\sigma})\prime\prime-{3\over2}
(\sigma)\prime\prime{e^{2\sigma}}):(w)\cr=
-{3\over4}\partial^2_w({1\over{(z-w)^2}})-{3\over2}
{1\over{(z-w)^4}}+...=-{6\over{(z-w)^4}}+...\cr
e^{-2\chi}(z)
:(-{1\over4}(e^{2\chi})\prime\prime\prime
+{3\over2}(\chi\prime\prime{e^{2\chi}})\prime):(w)
\cr=-{1\over4}\partial_w^3({1\over{(z-w)^4}})
+{3\over2}\partial_w({2\over{(z-w)^6}})+...
=-{{12}\over{(z-w)^7}}+...}}
and therefore
\eqn\grav{\eqalign{:\gamma^2b:(z){W_2}(w)\sim
{{2\times{-6}\times{-12}}\over{(z-w)^{4+7}}}ce^{2\phi}(w)
+...\cr
=144(z-w)^{-11}ce^{2\phi}(w)}}.
Then, using the above commutation relations and the fact that
$e^{2\phi}(z)e^{-5\phi}(w)\sim(z-w)^{10}e^{-3\phi}+...$
we find that the total commutator of the W-operator with the ghost
supercurrent
term of $Q_{brst}$ is given by:
\eqn\grav{\eqalign{\oint{dz}\gamma^2b(z)W(w)=
\oint{e^{2\phi-2\chi-\sigma}}(z)e^{-5\phi}(W_1-W_2)\psi_{m_1}...
\psi_{m_5}e^{ikX}(w)\cr=
\oint{{dz}\over{z-w}}(144ce^{-3\phi}\psi_{m_1}...\psi_{m_5}e^{ikX}
-144ce^{-3\phi}\psi_{m_1}...\psi_{m_5}e^{ikX})=0}}
Therefore the total commutator of $Q_{BRST}$ with $U_2$ is equal to zero
and the five-form is proven to be BRST non-trivial
(given its "transverse" five-dimensional propagation).
Equivalently, the BRST non-triviality of the 5-form
at picture $-3$ can be proven by considering the tree-point function
of $V_5^{(-3)}(k)$ with  $V_5^{(+1)}(p)$ and $V_B(q)$ where
$V_B$ is again a photonic vertex and k, p and q are
the corresponding momenta. For certainty, let us consider all the
operators in
the integrated form.
Since all these operators are proven to be BRST invariant,
the non-zero answer for the correlator will be a sufficient proof
for the BRST non-triviality. Since expressions for both photonic operator
in unintegrated form and  $V_5^{(-3)}(k)$ contain no fermionic ghosts,
only the first piece of $V_5^{(+1)}(k)$ without b and c fields will
contribute to the three-point correlator.
A simple computation gives:
\eqn\grav{\eqalign{<V_5^{(-3)}(k)V_5^{(+1)}(p)V_B(q)>\cr
=<e^{-3\phi}\psi_{m_1}...\psi_{m_5}e^{ikX}(z_1)
e^\phi\psi_{m_1}...\psi_{m_5}e^{ipX}(z_2)
e_m(q)(\partial{X^m}+i(q\psi)\psi^m)e^{iqX}(z_3)>\cr
=i(k^{\perp}e(q))^{\perp}{\times}{1\over{(z_1-z_2)(z_1-z_3)(z_2-z_3)}}}}
The coordinate dependence cancels out after being multiplied
by Koba-Nielsen's measure, so the final answer
for the correlator is given by just $i(k^{\perp}e(q))^{\perp}\neq{0}$
which proves the BRST non-triviality of the five-form if
$k=k^{\perp}$, i.e. is transverse to the $m_1...m_5$
directions which define the polarization of the five-form -
in complete agreement with our previous analysis. Technically,
 the transverse momentum dependense in the answer arises because
for the longitudinal ones the contribution
from the NSR fermionic part of $V_B(q)$ proportional to
$(q\psi)\psi^m$ cancels the one from the bosonic part and therefore
the total answer is zero;
while for the transversely oriented momenta
fermionic part of  $V_B(q)$
does not contribute while the bosonic part gives non-zero
contribution leading to the answer given above.
This concludes the analysis of the BRST properties of the
five-form vertex operators.

\centerline{\bf 4.Closed String Brane-like States and Effective Actions}
In this paper we have proven that
the brane-like vertex operators define new physical
states in superstring theory and are closely related
to the non-perturbative brane dynamics.
Closed string brane-like states can be constructed in a straightforward
way taking the vertices (1) as holomorphic and antiholomorphic parts
of closed string operators.
Then various polarizations and the way one contracts indices
in these operators would correspond to introducing various branes
and brane configurations.
As a simplest example consider a closed-string operator:
\eqn\grav{\eqalign{V_5=\int{d^2z}\epsilon_{a_1...a_6}\lambda(k^{\perp})
e^{-3\phi-{\bar\phi}}\psi_{a_1}...\psi_{a_5}
{\bar{\psi}}_{a_6}e^{ik^{\perp}X}(z,{\bar{z}})}}
where the 10d space-time index $m$ is split in the  $4+6$ way:
$m=(t,a);t=0,...,3,a=4,...,9$ and in this case
$ik^{\perp}X\equiv{k_t{X^t}}$.
As is easy to see, this closed string operator consists of
a holomorphic picture -3 five-form and antiholomorphic
photonic vertex  at picture -1 (the latter may of course be taken at any
picture). In this case, as is easy to see
by direct generalization of our previous analysis, BRST
invariance $+$ non-triviality condition require that the
momentum  $k^{\perp}$ propagates in $t$-directions only,
i.e. it is 4-dimansional;  $\lambda(k^{\perp})$ is a space-time scalar
field. The operator (97) has a $SO(1,3)\times{SO(6)}$ isometry,
i.e. the one of a D3-brane background. Therefore it is natural
to expect that inserting such an operator is equivalent
to introducing D3-branes in the theory and simulating
an effective geometry created by $D3-branes$.
Any choice of the scalar function $\lambda(k^{\perp})$
corresponds to particular distribution of the D3-branes in this picture.
In particular, it turns out ~{\myself} that
imposing certain constraints
on the form of this scalar function, namely,
choosing $\lambda(k^{\perp})\sim{1\over{(k^{\perp})^4}}$
creates a near-horizon geometry of D3-branes
playing a crucial role in AdS/CFT correspondence
~{\malda, \ampf,\wit}
 and can be used for
constructing the NSR superstring theory on $AdS_5\times{S^5}$
~{\kts, \myself}
To conclude this paper we would like to present the result
of computing the low-energy effective action corresponding
to sigma-model with the closed-string $V_5$-field,
i.e. the effective action for the space-time  $\lambda$-field
It turns out that this action has a DBI structure and the
 answer is given by:
\eqn\grav{\eqalign{
S_{eff}(\lambda)=\int{d^4}X{e^{-\varphi}}
{\sqrt{det(\eta_{st}+\partial_s\lambda\partial_t\lambda)}}}}
where the integration is taken over 4-dimensional subspace
defined by the $X^s$ coordinates $(s,t=0,...3)$
$\eta$ is Minkowski tensor and $\varphi$ is the dilaton field.
It is remarkable that the closed string sigma-model
with the operator (97) has a low-energy action with
open-string dilaton dependence (as it should be for D-branes).
Such an anomalous dilaton dependence in the effective action
(98) must be deeply related to the hidden RR-origin of the brane-like
states, observed in this paper.
The explanation for such an unusual dilaton coupling,
as well as detailed derivation of the effective action (98) will be
given in our next paper which is currently in progress.

\centerline{\bf Acknowledgements}
This work has been supported by the Academy of Finland
under the Project No. 163394.

\listrefs
\end